\documentclass[11pt,a4paper]{article}
\pdfoutput=1
\usepackage{jheppub}

\usepackage{amsmath}
\usepackage{epigraph}
\usepackage{hyperref}
\usepackage{color}
\usepackage{pifont}

\usepackage{bbm}
\usepackage{verbatim}   
\usepackage{subfigure}  
\usepackage{acronym}

\usepackage{amsfonts}
\usepackage{amssymb}
\usepackage{mathrsfs}
\usepackage{multirow}
 \usepackage{slashed}

\hypersetup{colorlinks=true,linkcolor=blue,citecolor=blue,urlcolor=blue}

\newcommand{\beq}{\begin{equation}}
\newcommand{\eeq}{\end{equation}}

\newcommand{\tev}{{\rm{TeV}}}
\newcommand{\gev}{{\rm{GeV}}}
\newcommand{\mev}{{\rm{MeV}}}

\def\slashchar#1{\ensuremath{                               %
   \setbox0=\hbox{${}#1{}$}       % set a box for #1 
   \dimen0=\wd0                                 % and get its size
   \setbox1=\hbox{/} \dimen1=\wd1               % get size of /
   \ifdim\dimen0>\dimen1                        % #1 is bigger
      \rlap{\hbox to \dimen0{\hfil/\hfil}}      % so center / in box
      {}#1{}                                    % and print #1
   \else                                        % / is bigger
      \rlap{\hbox to \dimen1{\hfil${}#1{}$\hfil}}   % so center #1
   \fi}} 

\begin{document}

\title{Electroweak relaxation from finite temperature} 
\author[a]{Edward Hardy,}
\emailAdd{ehardy@ictp.it}
\affiliation[a]{Abdus Salam International Centre for Theoretical Physics,
Strada Costiera 11, 34151, Trieste, Italy}

\abstract{
We study theories which naturally select a vacuum with parametrically small Electroweak Scale due to finite temperature effects in the early universe. In particular, there is a scalar with an approximate shift symmetry broken by a technically natural small coupling to the Higgs, and a temperature dependent potential. As the temperature of the universe drops, the scalar follows the minimum of its potential altering the Higgs mass squared parameter.  The scalar
 also has a periodic potential  with amplitude proportional to the Higgs expectation value, which traps it in a vacuum with a small Electroweak Scale. The 
required temperature dependence of the potential can occur through strong coupling effects in a hidden sector that are suppressed at high temperatures. Alternatively, it can be generated perturbatively from a one-loop thermal potential. In both cases, for the scalar to be displaced, a hidden sector must be reheated to temperatures significantly higher than the visible sector. However this does not violate observational constraints provided the hidden sector energy density is transferred to the visible sector without disrupting big bang nucleosynthesis. We also study how the mechanism can be 
implemented when the visible sector is completed to the Minimal Supersymmetric Standard Model at a high scale. Models with a UV cutoff of $10\,\rm{TeV}$ and no fields taking values over a range greater than $10^{12}\,\rm{GeV}$ are possible, although the scalar must have a range of order $10^8$ times the effective decay constant in the periodic part of its potential.}
\maketitle

%%%%%%%%%%%%%%%%%%
\section{Introduction}
%%%%%%%%%%%%%%%%%%
In a recent paper \cite{Graham:2015cka} models with a large number of metastable vacua were proposed in which, despite  the majority of the vacua having Higgs vacuum expectation values (VEV) close to the UV cutoff of the effective theory, a small Electroweak (EW) VEV is dynamically selected.\footnote{A similar mechanism was previously proposed to solve the cosmological constant problem  \cite{Abbott:1984qf}.} This relied on slow rolling of fields during inflation, an initial condition corresponding to no EW symmetry breaking, and crucially the boundary between zero Higgs VEV, $\left<h\right> = 0$, and $\left<h\right> >0$ being a special point in field space because of a periodic potential for an axion-like field proportional to $\left<h\right>$.

In this paper we consider models which select a vacuum with low EW scale in a similar way. However, instead of the important processes occurring during inflation, we utilise the properties of theories at finite temperature after reheating in the early universe.\footnote{The possibility that a particular vacuum may be preferred because of thermal effects has previously been considered in the context of meta-stable supersymmetry breaking \cite{Abel:2006cr,Craig:2006kx,Fischler:2006xh,Abel:2006my,Moreno:2009nk}, where it was shown that a supersymmetry breaking vacuum may be favoured over a deeper supersymmetry preserving vacuum. Finite temperature effects have also been studied in the context of the SM Higgs, which has a vacuum at large field values \cite{Espinosa:2007qp}.} Our model is a version of \cite{Graham:2015cka} in which, rather than rolling down a fixed potential, a scalar is always close to the minimum of an evolving potential, avoiding the need for a long period of inflation. In 
particular, there is an 
axion-like field $\phi$ 
with an approximate shift symmetry. The shift symmetry is explicitly broken by a small technically natural coupling to the Higgs and a  potential set by the same parameter. As the universe cools the potential for $\phi$ changes adiabatically, and we study models where this results in $\phi$ travelling a large distance in field space. Since the temperature of the universe changes on a timescale $1/H$, where $H$ is the Hubble parameter,  $\phi$ typically evolves over its field range on the same timescale (unlike the models of \cite{Graham:2015cka}).

The Lagrangian is chosen such that the evolution of $\phi$ reduces the Higgs mass squared parameter. We also assume that there is an interaction between $\phi$ and $h$ with a dependence  $\sim \Lambda_a^3 h  \cos\left(\phi /f_{eff} \right)$, where $\Lambda_a$ and $f_{eff}$ are mass scales. Provided the temperature in the visible sector is not too high, once the Higgs mass squared parameter becomes negative a Higgs VEV develops, and $\phi$ stops moving when the slope from the periodic potential is greater than that from the rest of the potential. This traps the Higgs and $\phi$ in a region of field space with a small EW VEV without tuning. 

The required temperature dependence of the $\phi$ potential can naturally arise from a hidden sector gauge group running into strong coupling, breaking the shift symmetry. The strength of the breaking generically depends on temperature \cite{Gross:1980br} (such behaviour is well known, e.g. for the QCD axion). As the temperature drops this contribution to the potential becomes  comparable to the explicit symmetry breaking part. If the potential is of a suitable form, for example similar to those generated by an anomalous coupling to a hidden sector gauge group or from axion monodromy \cite{Witten:1980sp,Kaloper:2011jz}, it can displace the minimum of the potential significantly.

A simple alternative  is that the temperature dependence occurs 
from  perturbative physics through a one-loop thermal potential. This can happen if $\phi$ has explicit shift-symmetry breaking couplings to states in a hidden sector in such a way that their masses depend on its VEV. However, as we discuss in Section \ref{sec:wc}, in this case a UV completion of the hidden sector must be specified. A suitable possibility is to make the hidden sector supersymmetric, with superpartner masses of order the visible sector UV cutoff or smaller. The resulting physics is similar to the strong coupling scenario, except that $\phi$ evolves due to the thermal potential becoming smaller as the temperature drops, rather than a sector becoming strongly coupled.

The details of the UV completion of the  Higgs sector can also affect the implementation of this type of mechanism. For consistency with the scenario where the hidden sector has (broken) supersymmetry, we show how the selection of a light EW scale can occur when the visible sector is UV completed to the Minimal Supersymmetric Model (MSSM) by superpartners with masses of order the cutoff. This is attractive from a top down perspective since the visible and hidden sectors then have supersymmetry broken at the same scale. While there are several alternatives, we focus on a theory where the visible sector effective $\mu$ parameter depends on $\phi$.

There is a significant model building constraint in both the strongly and weakly coupled hidden sector scenarios. For the minimum of the potential for $\phi$ to be displaced significantly a high temperature, of order the visible sector cutoff, is required. But, if $\phi$ has a high temperature it is not trapped by the barriers, which have height $\sim \Lambda_a^3\left<h\right>$.  To avoid this we consider models where the hidden and visible sectors are at different temperatures after inflation. For appropriate parameter choices $\phi$ is in thermal equilibrium with the visible sector, but not 
the high temperature hidden sector so is trapped successfully. Provided the hidden sector is such that its energy density is transferred to the visible sector at reasonably late times, heating the universe to above the scale of big bang nucleosynthesis (BBN) (but below $\sim 100\gev$ so that $\phi$ does not escape the barriers) this is compatible with cosmological observations.

A generic feature of the theories studied is that $\phi$ must travel over a large field range compared to the effective decay constant in the periodic part of its potential, although it does remain sub-Plankian. While worrying, for the purposes of this paper we simply assume that suitable models exist, and ignoring the UV completion of this part of the theory, the models considered are at least under control from an effective field theory perspective. Also, our models still have the `coincidence problem' of the original version \cite{Graham:2015cka}: a new gauge group is required to run into strong coupling close to the EW scale. We also find that to raise the visible sector cutoff significantly above the EW scale requires a visible sector temperature not far from the EW scale, which is a new but similar coincidence problem. Ultimately we find that the visible sector cutoff cannot be raised much above $10\,\tev$. However, this still corresponds to reducing the tuning by a factor of $10^4$ compared to the SM, 
and there may exist deformations that allow the cutoff to be raised further.

In what follows, we examine the details of these models when applied to the Standard Model (SM) Higgs, highlighting potential complications and constraints. In Section \ref{sec:sc} we consider models where the temperature dependence is generated from strong coupling effects. The case where the temperature dependence is generated from weak coupling is discussed in Section \ref{sec:wc}. In Section \ref{sec:susy} we show how the mechanism can work when the visible sector has broken supersymmetry, altering the details of the Higgs sector. Finally, in Section \ref{sec:dis} we consider possible extensions to the models studied here and conclude.

%%%%%%%%%%%%%%%%%%
\section{Strong Coupling Model} \label{sec:sc}
%%%%%%%%%%%%%%%%%%
As discussed, the Lagrangian contains an scalar $\phi$, with an approximate shift symmetry. This is explicitly broken by a dimensionless parameter $\epsilon$ that couples it to the Higgs and leads to a potential of the parametric form $V\left(\epsilon \phi \right)$ with mass parameters of order the visible sector cutoff $M$. Strong coupling in a hidden sector introduces further breaking and overall the zero temperature Lagrangian, ignoring coefficients of order $1$, is
\beq 
\mathcal{L} =  M^2 \left|h\right|^2 +  \lambda_h \left|h\right|^4 + {\rm{S.M.}} - \epsilon M \phi \left|h\right|^2 + V\left(\epsilon \phi\right) + \Lambda_a^3 h \cos\left(N \frac{\phi}{f} \right) - \Lambda_b^4 \left(\frac{\phi}{f}\right)^2  , \label{eq:La}
\eeq
where $\rm{S.M.}$ represents the other Standard Model interactions. 

The scale $\Lambda_b$ is assumed to be generated by strong coupling effects in a hidden sector. For example, an anomalous coupling of $\phi$ to a hidden sector gauge group with field strength $G_h$ of the form $\frac{\phi}{f} G_{h} \tilde{G_h}$ can lead to an interaction $\mathcal{L} \supset \Lambda_b^4 \cos \frac{\phi}{f}$ if the hidden sector gauge group runs into strong coupling. For $\frac{\phi}{f} \lesssim 1$ this reproduces the potential in Eq.\eqref{eq:La}. Alternatively, if $\phi$ couples anomalously to an ${\rm{SU}}\left(N\right)$ gauge theory in the large $N$ limit, the coupling Eq.\eqref{eq:La} is generated directly without expanding the potential around small $\phi$ \cite{Witten:1980sp,Kaloper:2011jz}. While, for definiteness, we restrict ourselves to this form of the potential, other powers or functional forms may be worth considering. Also, it will be seen that the mechanism can still work if this part of the potential has a positive sign provided the explicit symmetry breaking potential is 
chosen appropriately.

The explicit symmetry breaking part of the potential is under control as long as $\phi$ has a field range $\lesssim M/\epsilon$. In order that the energy density of the $\Lambda_b$ part of the potential remains $\lesssim \Lambda_b^4$ for this range of $\phi$,  we take
\beq \label{eq:frange}
f \gtrsim \frac{M}{\epsilon} .
\eeq

To avoid a large QCD $\theta$ parameter in conflict with observations, we assume that, similarly to models proposed in \cite{Graham:2015cka}, the $\Lambda_a$ term is not generated by QCD. Instead this coupling could arise from an additional gauge group, although this requires new EW charged fermions not far from the EW scale. Coexistence of this term with the $\Lambda_b$ requires $\phi$ to be coupled anomalously to two gauge groups, however this is not problematic. In such models it is hard to have $\Lambda_a \gtrsim 100 \,\gev$, and we take this as a constraint on our parameter space.  As discussed in  \cite{Graham:2015cka} collider limits do not currently rule out new EW charged states with the necessary masses, but it is interesting that they may be observable in the  future.  In these constructions $N$ is an anomaly coefficient, however we simply take it as a parameter that can have large values, without worrying about the model building implications. Alternatively, theories have been proposed in which 
the evolution of a second scalar allows periodic potentials, not associated to QCD, to be generated without new physics close to the EW 
scale  \cite{Espinosa:2015eda}. It may be possible to implement the features of this model in our framework by making the potential of both scalars temperature dependent. Although interesting, for simplicity we do not consider this in the present work.

The strong coupling physics generating $\Lambda_b$ is typically sensitive to the temperature and in particular we consider a dependence 
\beq \label{eq:tempde}
\Lambda_{b,T} \sim \begin{cases}
            \Lambda_{b} \left(\frac{\Lambda_{b}}{T_{hid}}\right)^n & T_{hid}> \Lambda_b \\[1em]   %%% 
            \Lambda_{b} & {\rm otherwise} .
        \end{cases}
\eeq
The power $n$ depends on the details of the hidden sector, and the 
resulting physics is not especially sensitive to its value. This form is well motivated, for example it appears for the QCD axion potential. More generally if the strong coupling physics is  suppressed, parametrically, by $\sim \exp\left(-1/g\left(\mu\right) \right)$, where $g$ is a gauge coupling, then since the renormalisation scale should be chosen  $\mu \sim T$ a power law dependence on temperature is obtained. The $\Lambda_a$ part of the potential generically has a similar dependence. 

The explicit symmetry breaking potential for $\phi$ is generically of order $M^4$, where $M$ is the UV cutoff of the visible sector of the theory. Therefore for a significant evolution of $\phi$, the sector generating the $\Lambda_b$ potential must be reheated to a high temperature $\gtrsim M$. 
Since $M$ is the UV cutoff only of the visible sector it is consistent for the hidden sector to have a much higher cutoff and so have such a temperature. At scales above $\Lambda_b$, this sector has only shift-symmetry preserving interactions with  $\phi$. Therefore, the higher cutoff does not lead to a larger potential for $\phi$ being radiatively generated provided $\Lambda_b \sim M$. However, the barriers trapping $\phi$ cannot be high enough to prevent thermal fluctuations exploring deeper minima if $\phi$ and the visible sector are also at high temperature. As a result, the visible sector, the $\Lambda_a$ part of the potential, and $\phi$ itself, are assumed to be reheated to a lower temperature and have a temperature $T_{vis}\lesssim 100 \,\gev$ while $\phi$ is evolving.\footnote{The possibility of different reheat temperatures in separate sectors has  been considered in e.g. \cite{Feng:2008mu}.} This ensures that the barriers trapping $\phi$ are not suppressed and thermal fluctuations do not allow $\
phi$ over the barriers.  We further require that $\phi$ remains out of thermal equilibrium with the high temperature sector, which will constrain the viable parameter 
space.

%%%%%%%%%%%%%%%%%%%%%%%%%%%%%%%%%%%%%%%%%%%%%%%%
%%%%%%%%%%%%%%%%%%%%%%%%%%%%%%%%%%%%%%%%%%%%%%%%

\subsection*{Evolution of $\phi$}

We first consider the theory immediately after reheating when a temperature $T_{hid}> M$ has been turned on in the $\Lambda_b$ sector. Because the temperature is high, the $\Lambda_b$ term in the Lagrangian is suppressed, and the potential for $\phi$ is set by the explicit symmetry breaking potential.\footnote{As well the zero temperature contribution, there will be a perturbative finite temperature potential. However, due to the small coupling $\epsilon$ and low temperature of the visible sector, this is negligibly small compared to the zero temperature piece.} As a simple example with representative behaviour we consider an explicit symmetry breaking potential
\beq \label{eq:phpot}
V\left(\epsilon\phi\right) =  \epsilon^2 M^2 \phi^2 + \epsilon^4 \phi^4 .
\eeq
The absence of a linear or cubic term is certainly not essential for the model, but keeps the formulas compact. Generally,  to ensure $\phi$ remains close to the minimum of the evolving potential, $V\left(\epsilon \phi\right)$ combined with the temperature dependent potential must be such that the minimum of the potential is a continuous function of temperature. The examples we consider satisfy this constraint, and it does not restrict the viable model space of more complicated potentials too severely.

After reheating, $\phi$ will begin at some point in its field range $\lesssim M/\epsilon$ and evolve towards the minimum of this potential  with a maximum velocity set by Hubble friction
\beq
\begin{aligned}
\dot{\phi}_{max} & \sim \frac{1}{H} \partial_{\phi} V .
\end{aligned}
\eeq
Therefore in one Hubble time, $\phi$ can evolve a distance
\beq
\begin{aligned}
 \Delta \phi &\sim \frac{\epsilon M^3 M_{pl}^2}{T_{hid}^4} ,
\end{aligned}
 \eeq
where $M_{pl}$ is the reduced Planck Mass. For the parameter ranges we consider $\phi$ quickly reaches its minimum, and thermalises with the visible sector. The  energy in $\phi$ and the visible sector can be redshifted away before $\phi_{min}$ starts evolving at $T_{hid} \sim M$ provided the hidden sector has a high enough temperature initially. In particular if the visible sector is reheated to a temperature $M$, for the visible sector temperature to have dropped below $100 \,\gev$ before the finite temperature evolution of $\phi_{min}$ begins we need 
\beq \label{eq:thidin}
T_{hid}|_{RH} \gtrsim \frac{M^2}{100\,\gev} ,
\eeq
where $T_{hid}|_{RH} $ is the temperature of the hidden sector after reheating. Alternatively, we can simply assume that the inflation and reheating dynamics are such that $\phi$ begins close to the minimum of the explicit symmetry breaking potential and the visible sector reheat temperature is $\lesssim 100\,\gev$.

\begin{figure}
\begin{center}
\includegraphics[width=0.99\textwidth]{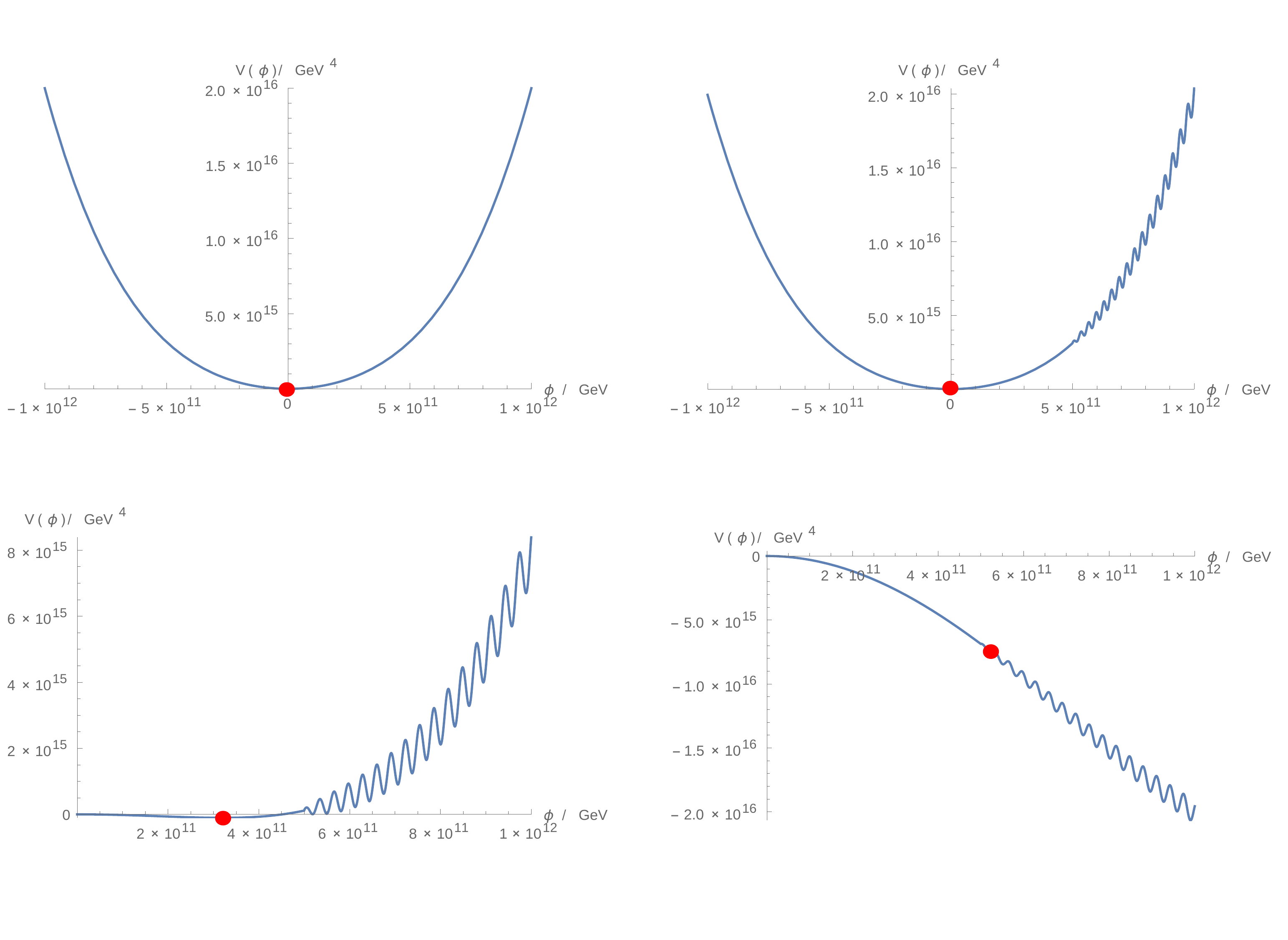}
\vspace{-30pt}
\caption{The potential for $\phi$ and the location of the field, indicated by a red dot, in the different temperatures regimes. \emph{Top left:} $T_{hid}> \Lambda_b$ and $T_{vis}> \Lambda_a$ when both strong coupling terms in the potential are strongly suppressed and the position of $\phi$ is set by the explicit shift symmetry breaking potential. \emph{Top right:} $T_{vis} < \Lambda_a$ and $T_{hid} > \Lambda_b$ when $\Lambda_a$ becomes strong but $\phi$ is yet to be displaced.  \emph{Bottom left:} $T_{vis} < \Lambda_a$ and $T_{hid} \sim \Lambda_b$ so $\phi_{min}$ and $\phi$ are moving. \emph{Bottom right:} $T_{vis} < \Lambda_a$ and $T_{hid} < \Lambda_b$ when $\phi$ is trapped in a local minimum of the potential with small Higgs VEV by the cosine part of the potential. In these plots $\epsilon = 10^{-8}$, $M=10^4 \,\gev$, and the period of the cosine has been greatly increased from realistic parameter ranges for visibility. For phenomenologically viable values of the cosine period, $\phi$ 
will not stop in the first 
local 
minimum it meets, but instead travel through many until the stopping condition Eq.\eqref{eq:cond2} is met.}
\label{fig:a}
\end{center}
\end{figure}

As the temperature drops further the $\Lambda_b$ part of the potential starts to become significant. Ignoring the Higgs contribution to the potential for the moment, the minimum as a function of the temperature is
\beq \label{eq:min}
\phi_{min} =  \frac{M}{\sqrt{2} \epsilon} \sqrt{\frac{\Lambda_{b,T}^4}{f^2 \epsilon^2 M^2}- 1 } ,
\eeq
when the argument of the square-root is positive, and $\phi_{min}=0$ otherwise. In this expression we have made a choice that the theory evolves towards a positive VEV. For $\phi$ to be displaced requires
\beq \label{eq:phrange}
\Lambda_{b}^2 > f \epsilon M .
\eeq
Combined with Eq.\eqref{eq:frange}, we need $\Lambda_b \gtrsim M$, and for simplicity we will typically consider parameters such that the inequalities are saturated, with $\Lambda_b \sim M $ and $f \sim M/\epsilon$. For the parameter ranges of interest $\phi$ remains in the range $\phi \lesssim M/\epsilon$ so the shift symmetry breaking potential is under control. The evolution of $\phi_{min}$ as the temperature changes is shown in figure \ref{fig:a} (which also includes the effect of the Higgs coupling, discussed shortly).

We also note that $\phi$ moves over most of its field range when $\Lambda_{b,T} \sim M $, at typical temperatures $T_{hid} \sim M$. The velocity of the minimum of the potential $\dot{\phi}_{min}$ when it is moving is of order
 \beq \label{eq:phspeed}
\dot{\phi}_{min} \sim \frac{\pi \sqrt{g_h}  M^3}{3\sqrt{10} \epsilon M_{pl} } ,
\eeq
where we have taken $n=1$ in Eq.\eqref{eq:tempde} for definiteness, and $g_h $ is the number of relativistic degrees of freedom in the high temperature hidden sector.

For the parameter regions of interest
\beq
\dot{\phi}_{min} \ll \frac{\partial_\phi V}{H} \sim \frac{\epsilon M^3 M_{pl}}{T^2} ,
\eeq
therefore the effect of Hubble friction can be neglected. The timescale over which $\phi_{min}$ travels a large distance in field space $\sim M/\epsilon$ is
\beq
\Delta t \sim \frac{M_{pl}}{\Lambda_b^2} ,
\eeq
which is parametrically the same as the Hubble time.

So far we have only considered the evolution of the minimum of the potential, but the interactions and thermalisation of $\phi$ itself are crucial. We will see later that for the most interesting points in parameter space the visible sector temperature satisfies $T_{vis} \sim \Lambda_a$. Under this assumption, the interactions of $\phi$ with the visible sector (and also with the physics that generates the periodic $\Lambda_a$ potential) occur at a rate $\Gamma_{vis}$ approximately given by
\beq
\begin{aligned} \label{eq:gvis}
\Gamma_{vis} &\sim T_{vis}^3 \left(\frac{N}{f}\right)^2 \\
\implies \frac{\Gamma_{vis}}{H} &\sim \frac{T_{vis}^3}{T_{hid}^2}  \left(\frac{N}{f}\right)^2  M_{pl} .
\end{aligned}
\eeq
For the parameter ranges of interest $\frac{\Gamma_{vis}}{H} \gg 1$, so $\phi$ remains in thermal equilibrium with the visible sector, on a timescale $\Delta t_{th} = \Gamma_{vis}^{-1}$.  In contrast the couplings to the hidden sector are much weaker and, since we are interested in times when $T_{hid} \sim M \sim \Lambda_b$, the rate of interaction is parametrically
\beq \label{eq:thermhs}
\begin{aligned}
\Gamma_{hid} &\sim T_{hid}^3 \left(\frac{1}{f}\right)^2 \\
\implies \frac{\Gamma_{hid}}{H} &\sim \frac{T_{hid} M_{pl}}{f^2} .
\end{aligned}
\eeq
In order that $\phi$ (and indirectly the visible sector) are not heated up we require 
\beq \label{eq:condthhid}
\frac{\Gamma_{hid}}{H} < 1. 
\eeq
If $T_{hid}|_{RH} \gg M$ so that $\phi$ can relax to its high temperature minimum after inflation, Eq.\eqref{eq:condthhid} must hold when $T_{hid} \sim M^2/ 100\,\gev$. At these temperatures the hidden sector is not strongly coupled, but there are still couplings of $\phi$ to, for example, hidden sector fermions (possibly suppressed by relatively small coupling constants, which would weaken this constraint). Alternatively, if we simply assume that $\phi$ begins from its high temperature minimum at $T_{hid} \sim M$, the condition is weakened and Eq.\eqref{eq:condthhid} must be satisfied at $T_{hid}\sim M$.

As its potential changes $\phi$ evolves following it. As a result it will oscillate around the minimum of its changing potential. Since the mass of $\phi$ is always much greater than the Hubble scale, these oscillations are fast and the evolution is close to adiabatic. Near its minimum, the time dependent potential for $\phi$ can be approximated
\beq \label{eq:tdp}
V\left(\phi\right) \sim \epsilon^2 M^2 \left(\phi - \dot{\phi}_{min} t \right)^2 .
\eeq
From its equation of motion it is seen that $\phi$ oscillates around the evolving minimum, with typical velocity $\dot{\phi}_{min}$ and typical amplitude $\dot{\phi}_{min}/\left( \epsilon  M\right)$. Therefore, provided
\beq 
\begin{aligned} 
\dot{\phi}_{min} &\lesssim T_{vis}^2 ,
\end{aligned}
\eeq
which implies
\beq
\begin{aligned} \label{eq:vcond}
\frac{M^3}{\epsilon M_{pl}} \lesssim \left(100 \,\gev\right)^2 ,
\end{aligned}
\eeq
the energy $\phi$ gains from the moving minimum will be safely less than its thermal energy from being in equilibrium with the visible sector, and $\phi$ can be trapped by the barriers from the periodic potential. However, this will be a severe constraint on the viable parameter regions. 

Importantly, $\phi$ does not gain energy $\sim M^4$ as it would if it was displaced a significant distance from its minimum in a fixed potential (neglecting Hubble friction); at any given time it is always close to the minimum of its potential. Therefore it is only sensitive to the fact that the potential changes, which affects it through Eq.\eqref{eq:tdp}, while the energy change due to $\Lambda_b$ becoming strong is purely associated to the hidden sector. Also the Hubble parameter is always such that quantum fluctuations are small, and $\dot{\phi_{min}}$ is small enough that Hubble friction plays no role in the evolution of $\phi$. The evolution of the theory is shown in figure \ref{fig:a} for some typical parameter choices.

%%%%%%%%%%%%%%%%%%%%%%%%%%%%%%%%%%%%%%%%%%%%%%%%%%%%%%%%%%%%%%%%%
%%%%%%%%%%%%%%%%%%%%%%%%%%%%%%%%%%%%%%%%%%%%%%%%%%%%%%%%%%%%%%%%%

\subsection*{Developing a Higgs VEV}

We now turn to the key process of the Higgs developing a VEV.  There is no guarantee that the potential and couplings of $\phi$ will be such that $m_h^2 =0$ at some point in the evolution of $\phi$. However, there is an order 1 probability that this will occur, and we simply regard it as a constraint on the form of the potential.\footnote{Introducing a small, natural, hierarchy between the parameter $\epsilon$ in $V\left(\epsilon \phi\right)$ and the $\epsilon$ that appears in the coupling to the Higgs would increase the field range of $\phi$ raising the probability of the appropriate behaviour occurring to $\sim 1/2$.}

Taking the Lagrangian of Eq.\eqref{eq:La} (with appropriate order $1$ factors) and the explicit symmetry breaking potential Eq.\eqref{eq:phpot} the Higgs mass squared parameter is
\beq
m_{h}^2 = \frac{1}{2} M^2 - \epsilon \phi M +  12 \lambda_h \left|h\right|^2 .
\eeq
If $\phi$ evolves as previously discussed, the Higgs mass squared parameter begins positive and decreases as the temperature approaches $\Lambda_b$. While $m_h^2 >0 $ (i.e. $\phi \lesssim \frac{M}{2\epsilon}$) the theory will stay on the locus $\left<h\right>=0$. As $\phi$ moves further the effective zero temperature Higgs mass squared parameter becomes negative. The Higgs interacts with the visible sector thermal bath strongly, so also experiences finite temperature effects. However, since the visible sector temperature is by assumption $\lesssim 100\,\gev$, this does not prevent a Higgs VEV developing once $m_{h}^2 \lesssim -\left(100 \,\gev \right)^2$. Also the low visible sector temperature means that the $\Lambda_a$ part of the potential is not suppressed by the temperature.

 As $\phi$ evolves following $\phi_{min}$ the Higgs VEV increases until at some point $\phi$ will be trapped by the $\Lambda_a$ part of the potential. This happens when the gradient from the periodic potential $\sim \Lambda_a^3 \left<h\right> N/f$ is steeper than that from the $\Lambda_b$ part of the potential for all lower temperatures. At later times $\phi$ remains in a local minimum, even though there are many deeper minima at larger field values, and the Higgs VEV is fixed. By choosing technically natural parameters  $\left<h\right> \sim 250 \,\gev$ can be obtained without tuning, as shown in figure \ref{fig:a}.

For $\phi$ to stop evolving when the Higgs has the correct VEV we need 
\beq \label{eq:cond2}
\begin{aligned}
\Lambda_a^3 \left<h\right> \frac{N}{f} &= \frac{\Lambda_b^4 \left<\phi\right>}{f^2}  \\
\implies \Lambda_a^3  \left<h\right> N  &\sim M^4 ,
\end{aligned}
\eeq
when $\left<h\right> \sim 250 \,\gev$. The gradient of the zero temperature part of the potential must also be smaller than that from the periodic potential, however this is automatically the case from the condition that $\phi_{min}$ is destabilised by $\Lambda_a$ given by Eq.\eqref{eq:phrange}. Eq.\eqref{eq:cond2} constrains the viable parameter space significantly because $\Lambda_a \lesssim 100 \,\gev$. It would be very interesting if models could be found allowing $\Lambda_a$ to be increased and therefore the cutoff raised. However, this is not straightforward  without either introducing tuning or radiatively generating a potential that destroys the dependence of the $\phi$ stopping position on the Higgs VEV \cite{Graham:2015cka,Espinosa:2015eda}.

We also need $\phi$ to have a small enough velocity that it does not roll over the barriers, which is the case provided Eq.\eqref{eq:vcond} is satisfied and the temperature of the visible sector is $T_{vis}^4 < \Lambda_a^3 \left<h\right> \lesssim \left(100\,\gev\right)^4$. 

To avoid tuning the vacua must be sufficiently close together, that is 
\beq
 \epsilon M f/N < \left(100\,\gev\right)^2 , \label{eq:cond3}
 \eeq
which is always satisfied for parameters satisfying the other constraints. Also, once the theory reaches a meta-stable vacuum  the rate of tunnelling to deeper ones must be tiny. The probability of this occurring can be found by considering the action of the so-called bounce solution $S_B$ \cite{Coleman:1977py,Coleman198802}. Although in general this is hard to compute, it can be roughly estimated by approximating the barrier as a potential for $\phi$ of 
\beq \label{eq:vba}
V \sim \lambda \left(\phi^2 - a^2\right)^2 - \frac{\epsilon}{a} \left(\phi - a \right),
\eeq
where $a$ sets the distance between the vacua, $\epsilon$ the energy difference between them, and $\lambda a^4$ the height of the barrier. If $\epsilon \ll \lambda a^4$, the ``thin-wall approximation'', the bounce action is of order \cite{Coleman:1977py}
\beq \label{eq:sbt}
S_B \sim \frac{\lambda^2 a^{12}}{\epsilon^3} ,
\eeq
and provided $S_B \gg 1$ the decay rate will be exponentially slow. While the thin-wall approximation is not always a very good one for the parameter ranges of interest here it at least gives a rough estimate of the rate. In particular taking the parameters in the potential Eq.\eqref{eq:vba} to fit our true potential,  the condition for vacuum decay to be exponentially slow, from Eq.\eqref{eq:sbt}, is
\beq \label{eq:tunt}
\left(\frac{f}{N}\right) \Lambda_a^6 \left<h\right>^2  \gg  \epsilon^3 M^9  .
\eeq
For the parameter ranges of interest this is always satisfied by several orders of magnitude, so tunnelling occurs on a timescale far longer than the age of the universe (and is sufficiently large that corrections from the exact form of the potential and the thin wall approximation are not expected to change this conclusion).

At some later time the energy density in the hidden sector must decay to the visible sector. The only constraints are that the visible sector is reheated to a temperature above $\sim \mev$ in order that BBN occurs, and below $\sim 100 \,\gev$ so that $\phi$ does not jump over the barriers and explore vacua with larger Higgs VEVS. The time for the hidden sector energy to decay to the visible sector can be made arbitrarily long, for example if the energy density in the hidden sector resides in states protected by an accidental symmetry. Since this depends on the details of the hidden sector, and is independent of the properties required for the rest of the mechanism, we do not consider this further in the present work.

Provided all of these conditions are satisfied, Eq.\eqref{eq:cond2} sets the point in field space at which $\phi$ will stop. 
 There are viable parameter regions that satisfy all of the conditions, although they all require a large value of $N$ and it is not possible to raise $M$ too far above the TeV scale. The dominant constraints are from $\phi$ not thermalising with the hidden sector Eq.\eqref{eq:condthhid}, $\phi$ not having too high a velocity Eq.\eqref{eq:vcond}, along with the condition for $\phi$ to stop Eq.\eqref{eq:cond2}, and the constraint on $\Lambda_b$ Eq.\eqref{eq:phrange}. A viable point in parameter space is
\beq \label{eq:parm2}
M= 10^4 \,\gev , \qquad \epsilon = 10^{-8}, \qquad f= 10^{12} \,\gev, \qquad N= 10^{8} ,\qquad \Lambda_a = 100 \,\gev ,
\eeq
with $\Lambda_b = M$, and a visible sector temperature of $100\,\gev$ when the hidden sector has temperature $M$.

%%%%%%%%%%%%%%%%%%%%%%%%%%%%%%%%%%%%%%%%%%%%%%%%%%%%%%
%%%%%%%%%%%%%%%%%%%%%%%%%%%%%%%%%%%%%%%%%%%%%%%%%%%%%%%%

\section{A Weakly Coupled Hidden Sector} \label{sec:wc}

While a strong coupled hidden sector is an interesting and plausible source for a potential with the required behaviour, it is desirable to have an example model in which the temperature dependence of the potential is fully calculable. In this section we consider a hidden sector that leads to the required temperature dependence through weak coupling effects. 

First we briefly review some standard results in thermal field theory  (for further details see for example \cite{Kapusta:2006pm}). At weak coupling, finite temperature effects can be introduced through the temperature dependent part of the one-loop potential
\beq
\begin{aligned}
V =  V_{0}\left(\phi,h\right) + & \frac{T^4}{2 \pi^2} \sum_b  \int_0^{\infty} dk \, k^2 \log\left(1  - \exp\left(-\sqrt{k^2 + \frac{m_b^2\left(\phi,h\right)}{T^2} } \right)\right) \\
& - \frac{T^4}{2 \pi^2} \sum_f  \int_0^{\infty} dk \, k^2 \log\left(1  + \exp\left(-\sqrt{k^2 + \frac{m_f^2\left(\phi,h\right)}{T^2} } \right)\right), \label{eq:Vt}
\end{aligned}
\eeq
where $b$ runs over the bosonic degrees of freedom and $f$ the fermionic degrees of freedom, and $m_i^2\left(\phi,h\right)$ is the mass squared parameter evaluated at a particular point in field space. At $T \gg m_i$ the potential can be expanded
\beq
V = V_{0}\left(\phi,h\right) +  \sum_b \left( -\frac{\pi^2 T^4}{90} + \frac{m_{b}^2\left(\phi,h\right) T^2}{24} \right) +  \sum_f \left( -\frac{7 \pi^2 T^4}{720}+ \frac{\left|m_{f}^2\left(\phi,h\right)\right| T^2}{48} \right)  . \label{eq:Vte}
\eeq
Hence, weakly coupled thermal field theory  favours fields having smaller, or in the bosonic case more negative, mass squared parameters. However, at $T<m_i$ heavy fields decouple from the theory as $V\sim e^{-M/T}$ and the expansion Eq.\eqref{eq:Vte} is invalid.

Suppose $\phi$ is coupled to states in a hidden sector such that its VEV determines their masses. This will generate a potential for $\phi$ at finite temperatures, which favours points in field space that minimise the hidden sector masses. As the temperature drops the finite temperature part of the potential becomes less important and the minimum of the potential will move towards its zero temperature value. As in the previous section, $\phi$ remains close to the minimum so also evolves. Also as before, the finite temperature potential has to displace $\phi$ from a zero temperature potential which has typical value $M^4$. Therefore, given that the hidden sector states cannot have masses greater than $T_{hid}$ without decoupling from the thermal bath, we again require $T_{hid} \sim M$ for $\phi$ to evolve. The temperature in the visible sector during this evolution must be $T_{vis} \lesssim 100\,\gev$.

Unlike the strongly coupled hidden sector previously discussed, the coupling of $\phi$ to the hidden sector explicitly breaks the shift symmetry. Therefore, we cannot simply take the cutoff of this sector to be much greater than $M$, because doing so would generate a  too large zero temperature potential for $\phi$. To consider hidden sector temperatures of order $M$ or higher we must specify a UV completion, and ensure this does not lead to a too large zero temperature potential for $\phi$.

One way to do this is to introduce supersymmetry to the $\phi$ and hidden sectors, with soft breaking scale in the hidden sector of order $\lesssim M$. Then the cutoff of the hidden sector can be much higher than $M$, and still only radiatively generate a zero temperature potential for $\phi$ of with typical magnitude $M^4$ due to non-renormalisation theorems \cite{Seiberg:1993vc}. In contrast, finite temperature breaks supersymmetry \cite{Boyanovsky:1983tu,Das:1978rx,Girardello:1980vv}. So there is a thermal potential with magnitude $\sim T_{hid}^2 \phi^2$, and the theory is under control, even when $T_{hid} \gg M$.

A simple realisation of this model is through an explicit shift symmetry breaking superpotential term
\beq \label{eq:sup}
\mathcal{L} = \int d^2 \theta \, \epsilon \Phi \Psi_1 \Psi_2 , 
\eeq
where $\Phi$ is a chiral multiplet containing $\phi$ as its $\theta=0$ component, and $\Psi_{1,2}$ are hidden sector chiral multiplets. We could also include soft symmetry breaking masses of order $M$ as well without changing anything important in the discussion. The term Eq.\eqref{eq:sup} leads to the components of $\Psi$ gaining $\phi$ dependent masses. Meanwhile $\phi$ itself is protected from gaining a large supersymmetry breaking mass by the shift symmetry, which we assume is respected by whatever interactions mediate supersymmetry breaking to the visible and hidden sectors.

The explicit shift symmetry breaking coupling Eq.\eqref{eq:sup} generates a finite temperature potential for $\phi$
\beq
V_T = \epsilon^2 \phi^2 T_{hid}^2 ,
\eeq
where we have dropped an order one factor for convenience. As in the strong coupling model, $\phi_{min}$ begins evolving from its high temperature value when the temperature of the hidden sector is $T_{hid} \sim M$. For the purposes of this paper it is sufficiently accurate to use the high temperature expansion of the potential (it can be checked that there are no significant quantitative differences if the full expression is used). As before we assume the hidden sector is reheated to $T_{hid}|_{RH} \gtrsim M^2/ 100\,\gev$. Then $\phi$ can relax to its high temperature minimum and the energy deposited in the visible sector redshift away sufficiently by the time $\phi_{min}$ starts to move.

For definiteness, we consider a zero temperature potential
\beq
V\left(\epsilon\phi\right) =  -\epsilon^2 M^2 \phi^2 + \epsilon^4 \phi^4 ,
\eeq
which has vacua at large field values. The evolution of the minimum with temperature is
\beq
\begin{aligned} \label{eq:phevo}
\phi_{min} &= \frac{M}{\sqrt{2} \epsilon} \sqrt{1-\frac{T_{hid}^2}{M^2}} ,
\end{aligned}
\eeq
when the argument of the square-root is positive and zero otherwise. We have made the choice that the theory moves towards the positive vacuum. The analysis of the model and constraints are very similar to that of Section \ref{sec:sc}, and we are brief here. The evolution of $\phi$ stops once it is trapped, which happens at a Higgs VEV of
\beq \label{eq:wcstop}
\Lambda_a^3 \left<h\right> \frac{N}{f} = \epsilon M^3 .
\eeq
The velocity of the minimum of the potential can be obtained from Eq.\eqref{eq:phevo} and is 
\beq
\dot{\phi}_{min} \sim \frac{M T_{hid}^2}{\epsilon M_{pl}} ,
\eeq
where $T_{hid} \sim M$ is the temperature when $\phi$ is evolving. The velocity of $\phi$ due to the evolving potential has the same parametric dependence as the previous model, and we still need the condition in Eq.\eqref{eq:vcond} to be satisfied. The rate of thermalisation of $\phi$ with the hidden sector is given by
\beq \label{eq:wcth}
\frac{\Gamma_{hid}}{H} \sim \frac{\epsilon^2 M_{pl}}{T_{hid}} ,
\eeq
for $T_{hid} \gtrsim M$ and suppressed otherwise. Thermalisation occurs through the shift symmetry violating interaction Eq.\eqref{eq:sup} leading to a different parametric dependence compared to the strong coupling model. As before, Eq.\eqref{eq:wcth} must be less than one at all times to avoid heating up $\phi$ and the visible sector. The thermalisation rate with the visible sector is again given by Eq.\eqref{eq:gvis}.  We must also require that the energy density in the hidden sector is transferred to the visible sector in such a way that the visible sector is reheated above the scale of BBN, but below $100 \,\gev$ so that $\phi$ does not jump over the barriers.

The constraints on the model are very similar to that of the previous section, except the finite temperature part of the potential no longer depends on $f$. Also, since the coupling to $\Psi$ is of order $\epsilon$ the potential of $\phi$ automatically changes at $T\sim M$ without having to choose a strong coupling scale $\Lambda_b \sim M$. Summarising we require: thermalisation with the visible sector is sufficiently fast  Eq.\eqref{eq:gvis}, thermalisation with the hidden sector is sufficiently slow Eq.\eqref{eq:wcth}, $\phi$ stops rolling in the correct place Eq.\eqref{eq:wcstop}, the rate of tunnelling is exponentially suppressed Eq.\eqref{eq:tunt}, and the temperature of the visible sector is small enough when the hidden sector temperature is $\sim M$. In particular, the parameters in Eq.\eqref{eq:parm2} are suitable here as well.

%%%%%%%%%%%%%%%%%%%%%%%%%%%%%%%%%%%%%%%%%%%%%%%%%%%%%%
%%%%%%%%%%%%%%%%%%%%%%%%%%%%%%%%%%%%%%%%%%%%%%%%%%%%%%%%

\section{UV Completing the Visible Sector with Supersymmetry} \label{sec:susy}
At its UV cutoff the visible sector must be completed to some other theory. While it is possible that the dynamics discussed are insensitive to the details of the high energy theory, this is not necessarily the case. As an example, we consider how vacuum selection can occur when the visible sector is completed to the MSSM by superpartners with masses of order $M \gg \tev$. The cutoff can then be raised all the way to the Planck Scale without a hierarchy problem. Even though the superpartners are heavy enough that there are no collider consequences at present (up to possible flavour and CP observables), the Higgs sector is still altered and we need to reconsider how the EW VEV changes during the evolution of $\phi$. This UV completion is particularly motivated in the supersymmetric version of the weak coupling model, since $\phi$ and the hidden sector are already supersymmetric with a soft breaking scale $\sim M$. 

As discussed in, e.g. \cite{Cheung:2005ba,Arvanitaki:2012ps}, being able to tune the theory to a small EW symmetry breaking vacuum by adjusting a single parameter is not always possible. In particular, EW symmetry breaking requires 
\beq \label{eq:ewsusy}
\left( \left|\mu\right|^2 + m_{Hu}^2 \right) \left( \left|\mu\right|^2 + m_{Hd}^2 \right) - B_{\mu}^2 < 0 .
\eeq
Here $m_{Hu}^2$ ($m_{Hd}^2$) is the soft mass squared of the up (down) type Higgs, $\mu$ is the SUSY preserving parameter in the superpotential $W \supset \mu H_u H_d$, and $B\mu$ is the parameter in the soft SUSY breaking potential $V_{soft} \supset  B\mu \, h_u h_d$, where $H_u$ ($H_d$) is the up (down) type Higgs chiral superfield with scalar components $h_u$ ($h_d$). Eq.\eqref{eq:ewsusy} ensures that the Higgs sector mass matrix has a negative eigenvalue, and for a small EW VEV this eigenvalue must be small. The parameters must also be such that there is not a run-away along a D-flat direction
\beq
2 B_{\mu} < 2 \left|\mu\right|^2 + m_{Hu}^2 + m_{Hd}^2 .
\eeq

Perhaps the simplest possibility for the required vacuum selection is for $\phi$ to modify the visible sector $\mu$ term from its bare value. For example as
\beq
\mathcal{L} \supset \int d^2 \theta \, \left( M H_u H_d- \epsilon \Phi H_u H_d \right),
\eeq
where $H_{u,d}$ are the two Higgs multiplets. In effect $\Phi$ couples like the singlet in the NMSSM, except does so very weakly \cite{Ellwanger:2009dp}.

We consider theories with $m_{Hu}^2 <0 $ and $m_{Hd}^2  >0 $ at the scale $M$. For simplicity we also assume $B_{\mu}$ is reasonably small compared to the other soft parameters, although successful vacuum selection is possible without this assumption. As $\phi_{min}$ evolves,  given by Eq.\eqref{eq:phevo}, the $\mu$ parameter decreases until Eq.\eqref{eq:ewsusy} is satisfied when an EW VEV develops. This increases as $\phi_{min}$ and $\phi$ move until $\phi$ becomes trapped in a local minimum as before. In particular, taking  $m_{Hu}^2 = -M^2/4$, and writing $ \mu_{eff}= M- \epsilon \phi$, the EW VEV $v$ is
\beq
\begin{aligned}
v & \simeq \frac{2}{\sqrt{g^2+g'^2}} \sqrt{- m_{Hu}^2 -  \left|\mu_{eff}\right|^2} \\
& \simeq \frac{2}{\sqrt{g^2+g'^2}} \sqrt{\epsilon \phi' M} ,
\end{aligned}
\eeq
or zero if the argument of the square-root is negative, where $\phi' = \phi - \frac{M}{2 \epsilon}$, and $g$, $g'$ are the ${\rm{SU}}\left(2\right)$, ${\rm{U}}\left(1\right)_Y$ gauge couplings respectively. Since  $\tan\beta$ is large  $v= \sqrt{v_{Hu}^2+v_{Hd}^2} \sim v_{Hu}$, where $v_{Hu}$ ($v_{Hd}$) is the VEV of the up-type (down-type) Higgs. 

Therefore, $\phi_{min}$ moving through the field range given by Eq.\eqref{eq:phevo} can indeed lead to successful EW symmetry breaking. As usual in supersymmetric theories the physical Higgs mass is not a free parameter of the theory and is instead fixed by the other parameters of the theory. This is why here the vacuum selection is phrased in terms of the Higgs VEV rather than its mass. For superpartners at the typical cutoffs of the theories considered, a physical Higgs mass $\sim 125\,\gev$ can be achieved from loop corrections in the MSSM (see for example \cite{Martin:1997ns,Vega:2015fna}).

%%%%%%%%%%%%%%%%%%%%%%%%%%%%%%%%%%%%%%%%%%%%%%%%%%%
%%%%%%%%%%%%%%%%%%%%%%%%%%%%%%%%%%%%%%%%%%%%%%%%%%%

\section{Discussion and Conclusions}  \label{sec:dis}
In this short paper we have considered a deformation of the models proposed in \cite{Graham:2015cka}. Instead of an axion-like state slowly rolling  down a potential during inflation, we have studied theories where it evolves due to its potential changing as the temperature of the universe drops. In particular, we considered the possibility that the temperature dependence arises from a strong coupling sector, the effects of which are expected to be suppressed at high temperatures, or through a weakly coupled hidden sector where the change happens due to a perturbative finite temperature potential. The behaviour of $\phi$ is reminiscent of the QCD axion $a$ \cite{Sikivie:2006ni}. If Pecci-Quinn 
symmetry is broken before inflation, the QCD axion takes a constant value $\left<a\right> = f_a \theta_0$ after inflation. This remains until the temperature is $\Lambda_{QCD}$, when the axion gets a potential and evolves. The crucial difference is that $\phi$ has an explicit shift symmetry breaking potential, whereas this is necessarily tiny for the QCD axion. Since this potential is present at all temperatures, $\phi$ evolves following the overall minimum of its potential, not starting very far from the minimum as for the QCD axion.

Of course much remains to be done. One shortcoming of the present work is a lack of an explicit example of a strong coupling sector with the required behaviour, although it is plausible they exist. Also, more problematically, the axion-like scalar is required to travel over many times its field range in the periodic part of the potential, which for typical models requires a very large anomaly coefficient. Obviously it is very interesting and important to see whether viable models where this occurs can actually be found. Further, there may well be other interesting classes of theories that lead to significant changes in the form of the potential at finite temperature.

Further, we have simply assumed the hidden and visible sectors are reheated to different temperatures. In order to claim the model is free of tuning such an initial condition must be naturally generated. One possibility is to just assume that inflaton couples dominantly to the hidden sector. For example an inflaton $\phi$ may decay through a coupling to hidden sector fermions $\psi$ through an interaction $\phi \psi \psi^{\dagger}$, but have no significant couplings to the visible sector. While not explaining the lack of e.g. a significant inflaton coupling to the visible sector Higgs squared, this is at least radiatively stable. Perhaps a more satisfactory alternative is for the inflaton to couple significantly to both the hidden and visible sectors, but dominantly reheat the former for dynamical reasons. One way this could occur is if the inflaton and hidden sector are such that the hidden sector is reheated by decays of the inflaton through broad resonances \cite{Kofman:1994rk,Kofman:1997yn}. This process 
can be exponentially efficient, so if 
it 
dominates the energy transfer from the inflaton, and the inflaton properties mean that similar processes are not as efficient at reheating the visible sector, large temperature differences can be generated. In a forthcoming paper it will be shown that such models can be constructed without tuning \cite{reheating}.

Throughout this paper we have worked with models where the periodic potential is generated through a new gauge group with new EW charged fermions close to the EW scale. Viable models where the scalar is the QCD axion would be very attractive. However, there are strong constraints on the QCD $\theta$ parameter that require the axion to have a very flat explicit shift symmetry breaking potential. Models along the lines of Section \ref{sec:wc} are somewhat promising in this regard because the finite temperature potential that is displacing $\phi$ disappears at low temperatures. However, the zero temperature potential must still be tiny, which in the absence of further 
dynamics requires the coupling to the Higgs to be similarly small. As a result, the field range required of $\phi$ is enormous, leading to concerns about the UV completion of the model and also $\phi$ having to evolve very fast and not being trapped by the barriers. It may however be possible to avoid these problems with further model building.

Unfortunately, the various constraints mean that in our models the visible sector cannot have a cutoff too far above the EW scale. Ultimately this means that for the cutoff to be raised significantly above the EW scale at all, the temperature of the visible sector must be close to the maximum value it can have without destroying the mechanism, i.e. the EW scale. This is a second coincidence problem (in addition to requiring the new scale $\Lambda_a$ to be close to the EW scale), forced on us by the relatively small viable parameter space. An interesting direction for future work is to attempt to find models with larger scales associated to the periodic potential. For the potentials we have considered $\sim \Lambda_a^3 h \cos \phi/f_{eff} $, raising $\Lambda_a$ is not possible since it is a new source of EW symmetry breaking. However, this restriction does not apply to interactions of the form $\Lambda_a^2 \left|h\right|^2 \cos \phi/f_{eff}$. Even in this case it is difficult to raise $\Lambda_a$ 
significantly 
above the EW scale, because doing so typically radiatively generates a too large potential for $\phi$ preventing the relaxation mechanism working. But it might be hoped that there is some way of evading this problem (for example, it has been addressed in \cite{Espinosa:2015eda}) allowing for higher visible sector UV cutoffs and new viable regions of parameter space.

If it is possible to raise the scale of the barriers  there is another deformation of the mechanism available. We have studied models where $\phi$ stops evolving once the slope of the period potential is greater than that of the rest of the potential. This occurs at visible sector temperature $< 100 \,\gev$ so that a Higgs VEV is present whenever its zero temperature mass squared parameter is negative. However the Higgs VEV could instead be selected by the temperature of the universe during the evolution of $\phi$. If the theory is such that $\phi$ is evolving at temperatures of $\sim 100\,\gev$, it gains an effective mass squared from thermal effects $\sim \left(100\,\gev\right)^2$. Therefore a Higgs VEV first develops when its zero temperature mass 
squared parameter is $\sim -\left(100\,\gev\right)^2$. If the barriers of the period potential are large, $\phi$ could then be trapped immediately. As the universe cooled the thermal Higgs mass would disappear, while $\phi$ remained trapped with a Higgs VEV of the correct size. This would remove the connection between the various parameters of the model, avoiding the coincidence problem that $\Lambda_a$ must be close to the EW scale.

Another possibility is to consider extending our models along the lines of \cite{Espinosa:2015eda}: the evolution of a second scalar field may allow potentials $\Lambda^2 h^2 \cos \phi/f_{eff}$ to be generated without new matter or interactions close to the EW scale. In principle the evolution of the second scalar field could occur as it tracks the temperature dependent minimum of a potential if there was a more complex hidden sector.

\acknowledgments
I am grateful to Asimina Arvanitaki, Savas Dimopoulos, Robert Lasenby,  Matthew McCullough, James Unwin, and Giovanni Villadoro for very useful discussions.

\bibliography{vacuum}
\bibliographystyle{JHEP}

\end{document}